\documentclass[%
 prd,
 reprint,
superscriptaddress,
preprintnumbers,
nofootinbib,
 amsmath,amssymb,
 aps,
]{revtex4-1}
\pdfoutput=1
\usepackage{hyperref}

\usepackage[textwidth=1.25cm,textsize=footnotesize,
]{todonotes}

\usepackage[utf8]{inputenc}
\newcommand{\be}{\begin{equation}}
\newcommand{\ee}{\end{equation}}
\newcommand{\ba}{\begin{eqnarray}}
\newcommand{\ea}{\end{eqnarray}}

\newcommand{\Cr}{\mathop\mathrm{Cr}}

\newcommand{\Ga}{\mathop\mathrm{Ga}}
\newcommand{\Ge}{\mathop\mathrm{Ge}}

\begin{document}

\preprint{INR-TH/2019-010}

\title{BEST potential in testing the eV-scale sterile neutrino explanation
 \\ of reactor antineutrino anomalies}

\author{Vladislav Barinov}
\email{barinov.vvl@gmail.com} 
\affiliation{Institute for Nuclear Research of the Russian Academy of Sciences,
  Moscow 117312, Russia}
\affiliation{Physics Department, Moscow State University, 
Vorobievy Gory, Moscow 119991, Russia}

\author{Vladimir Gavrin}
\email{gavrin@inr.ru}
\affiliation{Institute for Nuclear Research of the Russian Academy of Sciences,
  Moscow 117312, Russia}

\author{Valery Gorbachev}
\email{vvgor\_gfb1@mail.ru}
\affiliation{Institute for Nuclear Research of the Russian Academy of Sciences,
  Moscow 117312, Russia}

\author{Dmitry Gorbunov}
\email{gorby@ms2.inr.ac.ru}
\affiliation{Institute for Nuclear Research of the Russian Academy of Sciences,
  Moscow 117312, Russia}
\affiliation{Moscow Institute of Physics and Technology, 
  Dolgoprudny 141700, Russia}

\author{Tatiana Ibragimova}
\email{tvi@inr.ru}
\affiliation{Institute for Nuclear Research of the Russian Academy of Sciences,
  Moscow 117312, Russia}

%\date{}

\begin{abstract}
Baksan Experiment on Sterile Neutrino (BEST)
\cite{Gavrin:2010qj,Gavrin:2015boa,Barinov:2016znv} is presently at the stage
of production of the artificial neutrino source ${}^{51}\!\Cr$, the
gallium exposure will start in July and proceed for three
months. While aiming specifically at investigating the Gallium
neutrino anomaly (SAGE and GALLEX experiments) 
\cite{Abdurashitov:1998ne,Abdurashitov:2005tb,Kaether:2010ag}, BEST
can do more and it is tempting to estimate its ability in testing
sterile neutrino explanation of antineutrino (reactor) anomalies.
We observe a moderate sensitivity to the region in model
parameter space (sterile neutrino mass and mixing with active electron 
neutrino) outlined by the old reactor antineutrino anomaly
\cite{Mueller:2011nm,Huber:2011wv} and the 
best fit of DANSS experiment \cite{Alekseev:2018efk}, while
the Neutrino-4 favorite region \cite{Serebrov:2018vdw} 
falls right in the BEST ballpark. In particular, by analyzing SAGE+GALLEX and
Neutrino-4 $\chi^2$ distributions we find that Neutrino-4 results are fully
consistent with the Gallium anomaly, the significance of the combined
anomaly almost reaches 4$\sigma$ level. If the BEST confirms the
Neutrino-4 results, the joint analysis will indicate more than
5$\sigma$ evidence for the sterile neutrino of eV-scale mass.   
\end{abstract}

\maketitle

%%%%%%%%%%%%%%%%%%%%%%%%%%%%%%%%%%%%%%%%%%%%%%%%%%%%%%%%%%%%%%%%%
%%%%%%%%%%%%%%%%%%%%%%%%%%%%%%%%%%%%%%%%%%%%%%%%%%%%%%%%%%%%%%%%%

{\it 1. Introduction.} Neutrino sector of the Standard Model of particle physics (SM)
exhibits more and more puzzling aspects. Apart of neutrino
oscillations -- the only established phenomenon unambiguously
pointing at incompleteness of the SM -- there are so called neutrino
anomalies, for reviews see Refs.\,\cite{Abazajian:2012ys,Giunti:2019aiy}.
While the former require SM neutrinos to be
massive, the latter ask for departure from the standard pattern of the
three SM (active) neutrinos. The key issue is the new mass scale squared, $\Delta
m^2$, too high in comparison with 
the two mass squared differences extracted from the analysis of 
conventional neutrino oscillations\,\cite{Tanabashi:2018oca}. The
attractive solution (though its capability of solving all the
anomalies is questionable, e.g.\,\cite{Collin:2016rao}) is oscillations into new hypothetical light
neutrino, sterile with respect to the gauge interaction of
the SM. 

The anomalies wait for independent checks,
which when happenig often reveal results suffering from lack of
confidence or even announce new anomalies. Indeed, last year two new
experiments -- DANSS\,\cite{Alekseev:2016llm} and
Neutrino-4\,\cite{Serebrov:2015ros}, both dealing with short-base-line neutrino
oscillations -- have presented their results on searches for ${\cal
  O}$(eV) sterile neutrinos, which might be responsible for the
reactor antineutrino anomaly (RAA) \cite{Mueller:2011nm,Huber:2011wv}. Although the best fit
point in the plane (sterile neutrino mass squared $m^2_s=\Delta m^2$, mixing angle with
electron antineutrino $\theta$), referring to the reactor anomaly
(actually, to the joint Gallium-reactor anomaly, see below) 
has been excluded at 2$\sigma$ level, both experiments claim that
other (and different) regions in the model parameter space with
eV-scale sterile neutrinos are favored revealing
smallest $\chi^2$-values in the data analyses. The best fit point found by
DANSS is \cite{Alekseev:2018efk}
\begin{equation}
  \label{DANSS-results}
\Delta m^2 = 1.4\,\text{eV}^2\,,\;\;\;\;\;\sin^2 2\theta=0.05\,,
\end{equation}
while the Neutrino-4 collaboration claims 2.8$\sigma$ evidence for
oscillations of electron into sterile antineutrinos  
with parameters\,\cite{Serebrov:2018vdw}  
\begin{equation}
  \label{Neutrino-4-results}
\Delta m^2 = 7.34\,\text{eV}^2\,,\;\;\;\;\;\sin^2 2\theta=0.39\,.
\end{equation}
These two claims are new, and though some systematics issues possibly
relevant there are discussed in literature\,\cite{Danilov:2018dme}
(see also RENO hint\,\cite{RENO:2018pwo} 
 on time-dependent composition of the reactor fuel,
which might resolve RAA), 
they may be checked directly in the upcoming experiments on neutrino
oscillations.

%%%%%%%%%%%%%%%%%%%%%%%%%%%%%%%%%%%%%%%%%%%%%%%%%%%%%%%%%%%%%%%%%%%%%%%%%%%%%%%%%
\vskip 0.3cm 
{\it 2. Gallium anomaly.} 
In this paper we investigate BEST prospects in testing the sterile
neutrino explanation of these anomalies in the electron antineutrino
sector. The main purpose of BEST \cite{Gavrin:2010qj,Barinov:2016znv}
is to check directly the Gallium
anomaly\,\cite{Giunti:2010zu} -- deficits of electron neutrino events, observed by
SAGE\,\cite{Abdurashitov:1998ne,Abdurashitov:2005tb} and
GALLEX\,\cite{Kaether:2010ag} experiments in the neutrino capture reaction
\begin{equation}
  \nu_e + {}^{71}\!\Ga\rightarrow {\text e}^+ + {}^{71}\!\Ge
  \label{Gallium-Germanium}
\end{equation} 
at short 
distances from neutrino artificial sources. Both experiments have
performed two independent measurements with specially designed
artificial sources aiming at calibration of the detectors, which main
goal were measurements of the low-energy tail of the solar neutrino
flux. The combined results of the four calibrations can be explained
\cite{Laveder:2007zz} 
by oscillations into sterile (invisible) neutrinos with best fit parameters 
\cite{Barinov:2017ymq} 
\begin{equation}
  \label{Gallium-results}
\Delta m^2 = 2.5\,\text{eV}^2\,,\;\;\;\;\;\sin^2 2\theta=0.3\,.
\end{equation}

Although the Gallium anomaly happened in {\it neutrino sector}, within
the simplest sterile neutrino paradigm the model parameters ($\Delta
m^2$, $\theta$) must be the same provided by the
$CPT$-symmetry. Actually, the best fit values for the two anomalies
are close, and one can combine them in a joint anomaly, see
e.g.\,\cite{Giunti:2012tn}.

\begin{figure}[!htb]
  \centerline{
\includegraphics[width=\columnwidth]{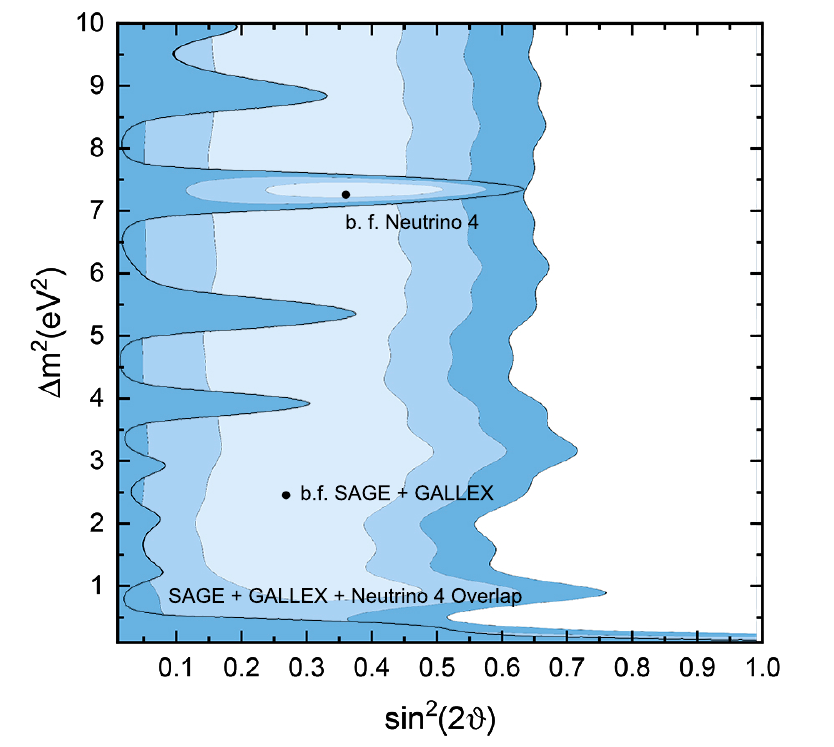}
  }
  \caption{Overlap of $\chi^2$ contours corresponding to the Gallium
    anomaly and Neutrino-4 results. Colors indicate regions of
    1$\sigma$,\ 2$\sigma$,\ 3$\sigma$ confidence levels (CL).
    Dots refer to the best fit points
    \eqref{Neutrino-4-results} and \eqref{Gallium-results}.
\label{Neutrino-4-bf-on-Gallium}
    }
\end{figure}

\begin{figure}[!htb]
  \centerline{
\includegraphics[width=\columnwidth]{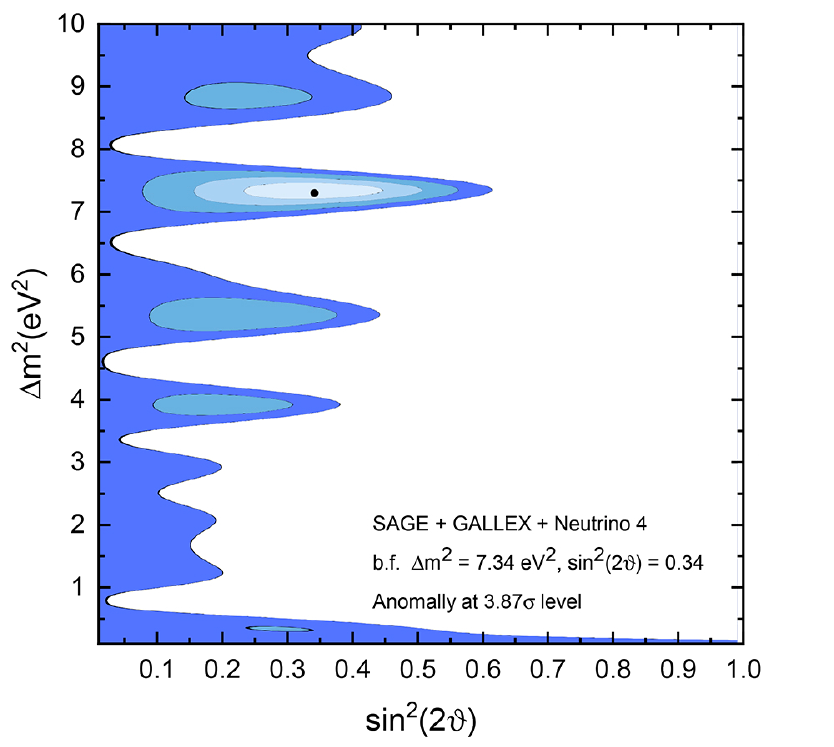}
  }
  \caption{
\label{Gallium-Neutrino-4-combined}
Regions favored by the combined Gallium and Neutrino-4 anomaly
at 1-4$\sigma$ CL.}
\end{figure}

Both experiments, DANSS and Neutrino-4
claim exclusion of the {\it joint anomaly} at 2$\sigma$ level
\cite{Alekseev:2018efk,Serebrov:2018vdw}, but
their sensitivity to each of the two anomalies differ. The reactor
antineutrino 
anomaly itself favors smaller mixing angle, than that of the joint
anomaly. It implies a lower signal and
higher statistics required for the 2$\sigma$ exclusion. On the
contrary, the Gallium anomaly prefers larger mixing angle, so that
{\it Neutrino-4 results \eqref{Neutrino-4-results} are fully consistent with the Gallium
  anomaly.} To illustrate this statement we present in
Fig.\,\ref{Neutrino-4-bf-on-Gallium} 
the contour plot of the 
$\chi^2$ distributions corresponding to both
anomalies\,\footnote{$\chi^2$ distribution of the Gallium anomaly is calculated in
  Ref.\,\cite{Barinov:2017ymq}; we thank the Neutrino-4 collaboration
  for sharing its \textbf{$\chi^2$} data analyzed in
  Ref.\,\cite{Serebrov:2018vdw}. Note that $\chi^2$ contours 
in our Fig.\,\ref{Neutrino-4-bf-on-Gallium} are a little bit different
from the contours on plots of Ref.\,\cite{Serebrov:2018vdw}, because 
the Neutrino-4 collaboration provided us with the updated $\chi^2$
distribution corrected for the systematics used in the
concluding part of the paper \cite{Serebrov:2018vdw} to estimate 
significance of the Neutrino-4 anomaly.} 
One observes that the Neutrino-4 and
Gallium 1$\sigma$ contours are widely overlapped, the best fit point of
Neutrino-4 is within 1$\sigma$ contour of the Gallium anomaly. 
To further confirm the consistency of the two anomalies, we follow
Ref.\,\cite{Barinov:2016znv,Barinov:2017ymq} and present in
Fig.\,\ref{Gallium-Neutrino-4-combined}
the likelihood for the joint
analysis of the Gallium and Neutrino-4 results, assuming one and the
same sterile neutrino to be responsible for both anomalies. The
significance of the joint anomaly almost reaches 4$\sigma$ and the best fit
point is close to that of Neutrino-4.

%%%%%%%%%%%%%%%%%%%%%%%%%%%%%%%%%%%%%%%%%%%%%%%%%%%%%%%%%%%%%%%%%%%%%%%%%%%%%%
\vskip 0.3cm 
{\it 3. BEST present status and prospects.}  
To check the Gallium anomaly BEST will use the artificial neutrino
source ${}^{51}\!\Cr$ of 3\,MCi to be placed in the center of a
spherical vessel filled with liquid gallium metal target and placed, in turn, in
the middle of a cylindrical vessel also filled with gallium metal
target \cite{Gavrin:2015boa}. Thus, the 
gallium target in both vessels will be exposed to neutrino flux, and because of
the reaction \eqref{Gallium-Germanium} 
the ${}^{71}\!\Ge$ atoms will appear via neutrino capture.
Then these atoms will be extracted and counted 
for each gallium target
providing direct measurements of the electron neutrino
flux averaged over 
each gallium target volume.
The activity of the source will be measured by calorimetry \cite{Kozlova:2018boa} and other methods \cite{Gorbachev:2015boa} with accuracy exceeding 1\%.
Since the neutrino capture rate is the same in both gallium targets,
the  
extractions from both vessels will be used independently to measure the neutrino flux.
 If the Gallium
anomaly is really the first evidence for sterile neutrinos, BEST will
observe deficits of events \eqref{Gallium-Germanium} in each vessel;
the particular numbers depend on the sterile neutrino parameters. The
BEST geometry is chosen in order to optimize its sensitivity and make
it the highest for the model parameters close to the best fit point of the
Gallium anomaly \eqref{Gallium-results}.

At the first stages of experiment the vessel for gallium has been
constructed and the techniques of filling it with gallium and emptying
it have been developed. The gallium has been exposed to the solar
neutrinos and the emerged germanium nuclei have been extracted
following the same procedure that will be used for BEST, revealing
results fully consistent with predictions of solar neutrino
physics. 
Meanwhile, two independent methods of
high-precision measurement of the power of BEST neutrino artificial
source --  ${}^{51}\!\Cr$ of 3\,MCi -- have been developed 
\cite{Kozlova:2018boa,Gorbachev:2015boa}. 

The high-power artificial source is the most expensive part of
BEST and the experiment has been approved and received the full financial
support only a year and half ago. 
Since then several key milestones of
the project have been passed. Presently the chromium source is
irradiating at SM-3 reactor in Dimitrovgrad to reach the required
intensity. The procedure will be completed by July and the source will
be transported to the Baksan Neutrino Observatory of INR RAS. There it
will be placed inside the specially  designed vessel and radiate
gallium for three months. During this period there will be several
extractions of germanium
nuclei, which are produced in the process \eqref{Gallium-Germanium}.
The expected sensitivity to the sterile neutrino model
explaining the Gallium anomaly has been estimated in
Ref.\,\cite{Barinov:2016znv} and further refined in
Ref.\,\cite{Barinov:2017ymq}.

%%%%%%%%%%%%%%%%%%%%%%%%%%%%%%%%%%%%%%%%%%%%%%%%%%%%%%%%%%%%%%%%%%%%

\begin{figure}[!htb]
  \centerline{
\includegraphics[width=\columnwidth]{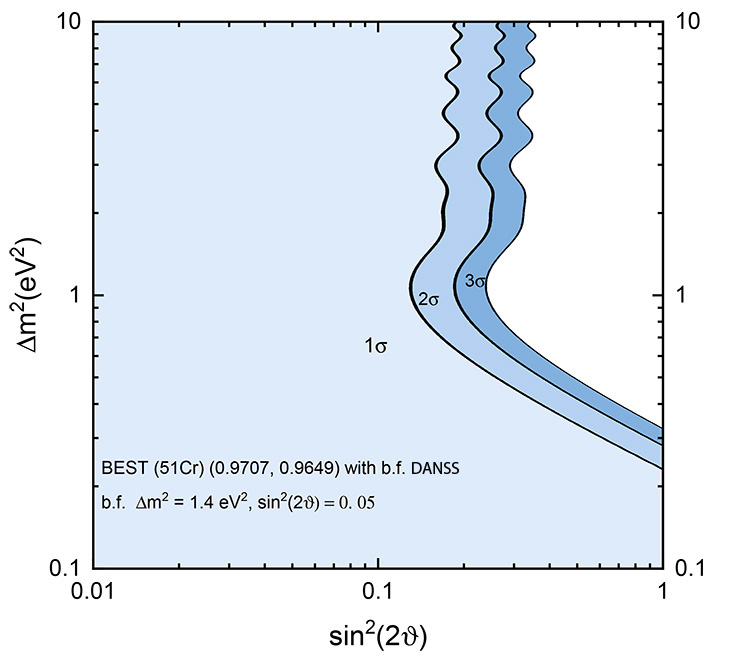}
  }
  \caption{Parameter space region to be favored by future BEST results
    at 1-3$\sigma$ CL, if it
    confirms the DANSS best fit \eqref{DANSS-results}. 
\label{BEST-DANSS}
    }
\end{figure}

\begin{figure}[!htb]
  \centerline{
\includegraphics[width=\columnwidth]{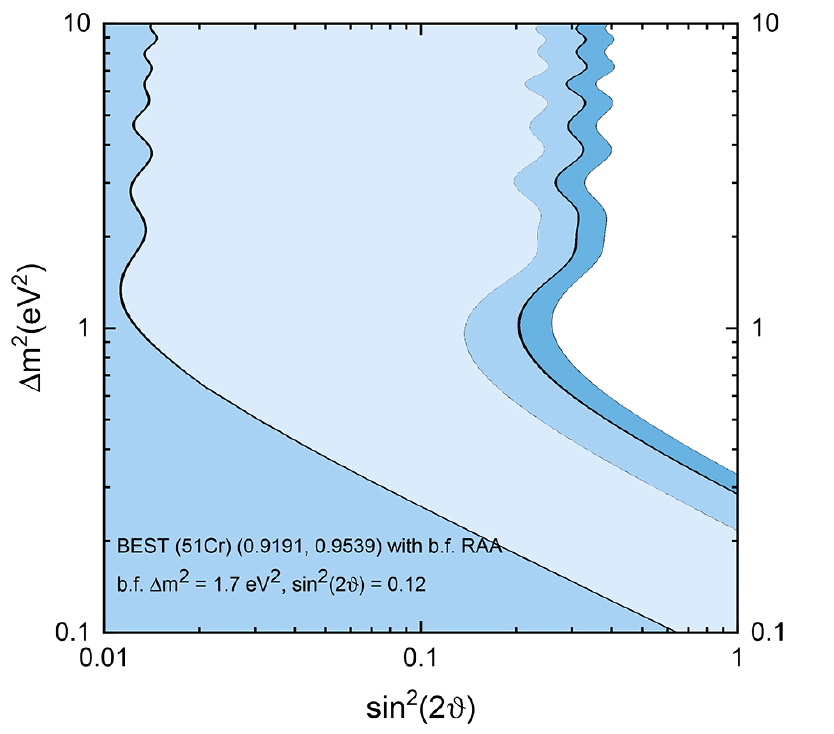}
  }
  \caption{The region in model parameter space to be favored by future
    BEST results at 1-3$\sigma$ CL if it
    confirms the reactor antineutrino anomaly (RAA) \eqref{reactor-results}.
    \label{BEST-reactor}
    }
\end{figure}

\begin{figure}[!htb]
  \centerline{
\includegraphics[width=\columnwidth]{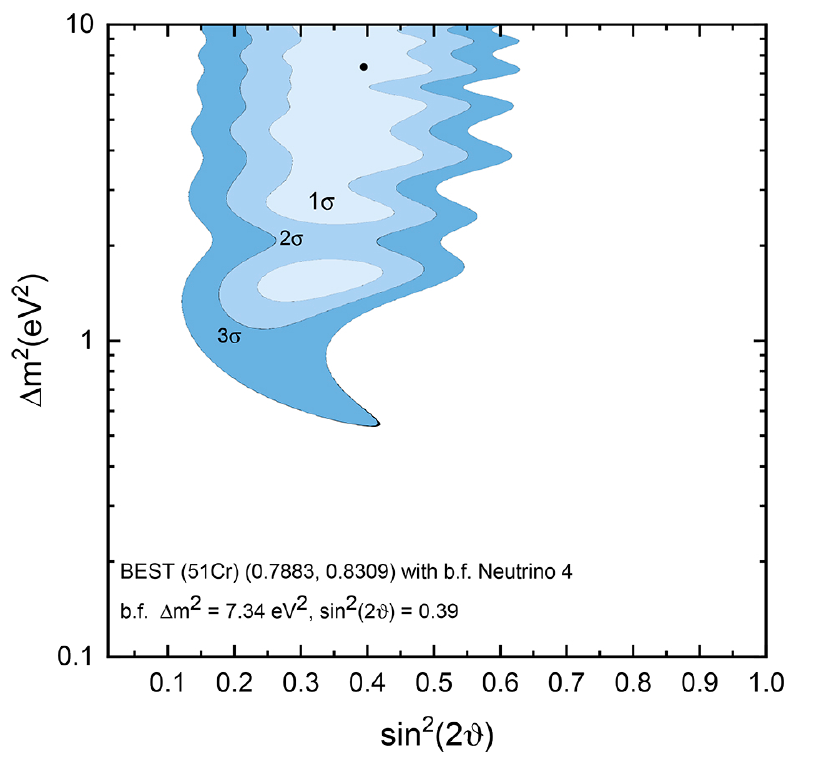}
  }
  \caption{The region in sterile neutrino model parameter space to be
    favored by future BEST results at 1-3$\sigma$ CL if it
    confirms the Neutrino-4 anomaly \eqref{Neutrino-4-results}.
    \label{BEST-Neutrino-4}
    }
\end{figure}

\begin{figure}[!htb]
  \centerline{
\includegraphics[width=\columnwidth]{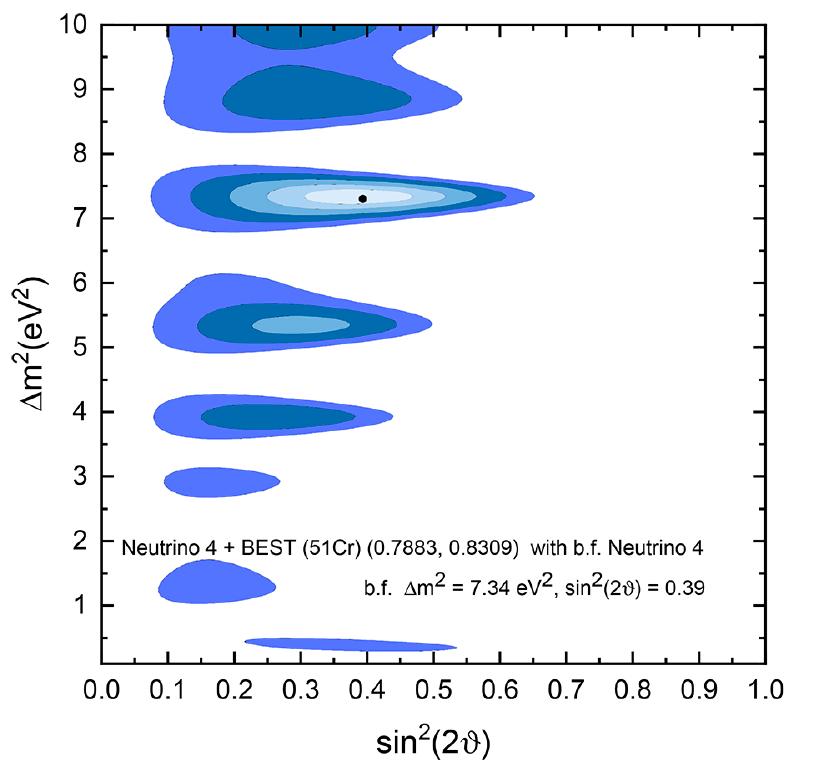}
  }
  \caption{The region of sterile neutrino parameters to be favored at
    1-5$\sigma$ CL by joint analysis of
    Neutrino-4 and future BEST results if the latter
    confirms the former \eqref{Neutrino-4-results}.
    \label{joint-5}
    }
\end{figure}

\vskip 0.3cm
{\it 4. Testing the recent anomalies at BEST.} 
Given the optimization based on the Gallium anomaly best fit
\eqref{Gallium-results} discussed
above, BEST exhibits higher sensitivity to the Neutrino-4 favored
region\,\eqref{Neutrino-4-results} than to that of 
DANSS\,\eqref{DANSS-results} or that of the original reactor
antineutrino anomaly,
which best fit value (in the fixed flux case \cite{Dentler:2017tkw}, \cite{Giunti:2012tn}) is
\begin{equation}
  \label{reactor-results}
\Delta m^2 = 1.7\,\text{eV}^2\,,\;\;\;\;\;\sin^2 2\theta=0.12\,.
\end{equation}
To illustrate the BEST abilities we preset in Fig.\,\ref{BEST-DANSS}
and
Fig.\,\ref{BEST-reactor}
the regions to be preferred by BEST should its
future results confirm the DANSS and reactor anomaly best fit values,
respectively. One observes, that BEST results could contribute to the
significances of corresponding anomalies, but rather modestly. On the
contrary, if future BEST results confirm the Neutrino-4 claim, it will
imply stronger than 3$\sigma$ confirmation, see
Fig.\,\ref{BEST-Neutrino-4}.
In that case, if combined with Neutrino-4
data, the joint anomaly will exceed 5$\sigma$ level, see
Fig.\,\ref{joint-5},
typically
accepted as a discovery condition in particle physics. 

So far we have considered the BEST ability in confirming the anomalies
and hence discovering the new physics. This is the most attractive
situation, however it is not guaranteed, and all the anomalies can
disappear with results of upcoming experiments. In particular, the 
analysis of Ref.\,\cite{Barinov:2017ymq} ensures that if BEST confirms the standard
three-neutrino oscillation pattern, see Fig.\,4 there, the Neutrino-4 anomaly will be
excluded at more than 3$\sigma$ level. As BEST sensitivities to the
reactor anomaly and DANSS best fit point are worse, the corresponding
exclusion power there is too low to change their status.      

%%%%%%%%%%%%%%%%%%%%%%%%%%%%%%%%%%%%%%%%%%%%%%%%%%%%%%%%%%%%%%%%%%%%

\vskip 0.3cm
{\it 5. Conclusions.} 
To summarize, we analyze the sensitivity of BEST to the regions in
the sterile neutrino model parameter space capable of explaining 
anomalies in electron antineutrino oscillation experiments: the (old) reactor
antineutrino anomaly and the recent results of DANSS and Neutrino-4 experiments.  

\vskip 0.3cm
We thank M.\,Danilov, D.\,Svirida and A.\,Serebrov for valuable
discussions and acknowledge sharing the Neutrino-4 likelihood
distribution by R.\,Samoilov and A.\,Serebrov.  
The work was performed using the scientific equipment of UNU GGNT BNO INR RAS 
with partial financial support of the Ministry of education and science of the Russian Federation: agreement 14.619.21.0009, unique identifier of the project RFMEFI61917X0009.
The estimate of BEST sensitivity to the reactor and Neutrino-4
anomalies was supported by the RSF grant 17-22-01547.

%%%%%%%%%%%%%%%%%%%%%%%%%%%%%%%%%%%%%%%%%%%%%%%%%%%%%%%%%%%%%%%%%%%
\bibliographystyle{apsrev4-1}
\bibliography{Best4neutrino}

\end{document}